\documentclass[aps,prl,preprint,superscriptaddress]{revtex4-1}

\usepackage{graphicx}
\usepackage{dcolumn}
\usepackage{bm}

\begin{document}


\title{Superconductivity of platinum hydride}


\author{Takahiro Matsuoka}
\email[]{tmatsuoka6@gmail.com/tmatsuok@utk.edu}
\affiliation{Department of Electrical, Electronic and Computer Engineering, Gifu University, Gifu, Japan}

\author{Masahiro Hishida}
\affiliation{Department of Materials Science and Technology, Gifu University, Gifu, Japan}

\author{Keiji Kuno}
\affiliation{Environmental and Renewable Energy Systems Division, Gifu University, Gifu, Japan}

\author{Naohisa Hirao}
\affiliation{Japan Synchrotron Radiation Research Institute, Hyogo, Japan}

\author{Yasuo Ohishi}
\affiliation{Japan Synchrotron Radiation Research Institute, Hyogo, Japan}

\author{Shigeo Sasaki}
\affiliation{Department of Electrical, Electronic and Computer Engineering, Gifu University, Gifu, Japan}

\author{Kazushi Takahama}
\affiliation{Center for Science and Technology under Extreme Conditions, Graduate School of Engineering Science, Osaka University, Japan}

\author{Katsuya Shimizu}
\affiliation{Center for Science and Technology under Extreme Conditions, Graduate School of Engineering Science, Osaka University, Japan}


\date{\today}

\begin{abstract}
We report the ac magnetic susceptibility, electrical resistance, and X-ray diffraction measurements of platinum hydride (PtH\textit{$_{x}$}) in diamond anvil cells, which reveal its superconducting transition. 
At 32 GPa, when PtH\textit{$_{x}$} is in a \textit{P}6$_{3}$/\textit{mmc} structure, PtH\textit{$_{x}$}  exhibits superconducting transition at 6.7 K and superconducting transition temperature (\textit{T}$_{c}$) decreases with pressure to 4.8 K at 36 GPa. The observed \textit{T}$_{c}$ is higher than that of powdered Pt by more than three orders of magnitude. It is suggested that hydrides of noble metals have higher \textit{T}$_{c}$ than the elements. 
\end{abstract}

\pacs{}

\maketitle

Pt, a well-known chemically inert noble metal, exhibits strong electron--phonon (\textit{e}--\textit{p}) coupling, which is favorable for superconductivity \cite{H.J.ALBERTandL.R.RUBIN1971, Herrmannsdorfer1996, Superconductivitybook}.
In addition, its mass magnetic susceptibility (1.21$\times$ 10$^{-8}$ (m$^{3}$/kg in SI unit)) is one of the largest ones among metals that do not show magnetic ordering \cite{H.J.ALBERTandL.R.RUBIN1971, Herrmannsdorfer1996}.
Consequently, the 5\textit{d} conduction electrons of Pt show strong spin fluctuations on short length and time scales, which tend to suppress the superconducting transition \cite{Superconductivitybook}. 
This large paramagnetic Pauli susceptibility also enhances the effective moments of magnetic 3\textit{d} impurities (e.g., Fe, Mn, and Co) because of the polarization of the neighboring 5\textit{d} conduction electrons of the host metal \cite{Mydosh}. 
So far, no study has reported the superconducting transition of bulk Pt, either at ambient pressure or high pressures reaching 300 GPa.
Having these backgrounds, it is natural to consider the development of a new superconductor that has a superconducting transition temperature (\textit{T}$_{c}$) as high as possible from this noble metal. 
K\"{o}nig \textit{et al}. found that powder Pt with an average grain size of approximately 2 $\mu$m showed a magnetic behavior very different from that of a bulk material; powder Pt shows a much weaker temperature dependence on dynamic susceptibility in the mK and $\mu$K regions \cite{Konig1999}.  
Then, the superconducting transition was observed at \textit{T}$_{c}$ = 0.62 - 1.38 mK \cite{Konig1999}. 
The lattice softening and enhanced \textit{e}-\textit{p} coupling, resulting from a large surface-to-volume ratio, have been suggested to play roles in the superconductivity of powdered Pt \cite{Konig1999,Schindler2000}. 

To manipulate the electronic states of metals, hydrogenation is sometimes an effective means. 
A good example is palladium hydride (PdH$_{\textit{x}}$). 
Pd has a high mass magnetic susceptibility of +6.57$\times$10$^{-8}$ (m$^{3}$/kg in SI unit) and is paramagnetic \cite{H.J.ALBERTandL.R.RUBIN1971}. 
The superconductivity of Pd has not been detected down to 2$\times$10$^{-4}$ K at 1 bar \cite{H.J.ALBERTandL.R.RUBIN1971}. 
When Pd forms a hydride PdH\textit{$_{x}$}, which has a NaCl-type crystal structure, the \textit{s} electrons from the H atoms occupy the 1\textit{s}-4\textit{d} bonding band and the unoccupied \textit{d}-band of Pd \cite{Papaconstantopoulos1980,Burger1981}. 
This Pd 4\textit{d} and H 1\textit{s} electron hybridization results in the reduction in electron density of states (e-DOS) at a Fermi level (\textit{E}$_{F}$), leading to the loss of paramagnetic susceptibility. 
PdH\textit{$_{x}$} becomes nonmagnetic with hydrogen composition \textit{x} = H/Pd greater than 0.62 \cite{Hara2006, Miller1975}. 
At the same time, the coupling between conduction electrons and the soft-optic mode, which is related to local hydrogen vibrations, is enhanced \cite{Papaconstantopoulos1980,Burger1981}. 
As a result, the superconductivity of PdH\textit{$_{x}$} appears at 1 K for \textit{x} $\sim$ 0.8, and \textit{T}$_{c}$ increases to 9 K as \textit{x} increases to 1 \cite{Skoskiewicz1972,Stritzker1972}.
The electronic band structures of elemental Pt and Pd share a similar feature: the broad dispersion bands of the \textit{d} electrons crossing the \textit{E}$_{F}$ and the large e-DOS at \textit{E}$_{F}$  \cite{Papaconstantopoulos2015}. 
In PtH, the hybridization between Pt 5\textit{d} and H 1\textit{s} \cite{Papaconstantopoulos1980,Zhang2012} and the resulting strengthened \textit{e}-\textit{p} coupling are anticipated \cite{Kim2011,Scheler2011,Zhou2011,Papaconstantopoulos1980,Zhang2012}. 
In addition, this hybridization can lead to a reduction in e-DOS at \textit{E}$_{F}$ and the loss of magnetic susceptibility. 
Theoretical calculations have predicted the \textit{T}$_{c}$ at the temperatures of 10-25 K  \cite{Kim2011,Scheler2011,Zhou2011,Papaconstantopoulos1980,Zhang2012}. 
Although Pt does not react with H$_{2}$ at ambient pressure, recent X-ray diffraction (XRD) experiments have revealed the formation of PdH\textit{$_{x}$} (\textit{x} $\sim$1) in high-pressure H$_{2}$ above 27 GPa at room temperature \cite{Scheler2011,Hirao2008_1}. 
This compound exhibits two phases, PtH-I and PtH-II, coexisting up to 42 GPa at room temperature, above which the single phase of PtH-II exists \cite{Scheler2011,Hirao2008_1}. 
The crystal structure of PtH-I is complicated, and \textit{P}321 is suggested, in which H atoms occupy tetrahedral interstitial sites \cite{Scheler2011}. 
In PtH-II, Pt atoms form a hexagonal close-packed lattice (\textit{P}6$_{3}$/\textit{mmc}), while H atoms are predicted to be in the octahedral interstitial sites \cite{Scheler2011,Hirao2008_1}. 
Note that the \textit{T}$_{c}$ predicted by \textit{ab-initio} calculations based on the harmonic approximation of phonon spectra lies in the range of 10-25 K \cite{Kim2011,Scheler2011,Zhou2011,Zhang2012}. 
On the other hand, the \textit{T}$_{c}$ is suppressed below 0.4 K when anharmonic effects, which often have a considerable degree of importance in the physical properties of hydrogen-rich compounds, are included in the calculation \cite{Errea2014}.
Thus, the superconductivity of PtH$_{x}$ is a fundamental question not only in the pursuit of higher \textit{T}$_{c}$ but also in the theoretical designing of hydrogen-related superconductors.

In this paper, we report the measurements of the ac magnetic susceptibility, electrical resistance, and XRD of PtH$_{x}$ in diamond anvil cells (DACs).
Above 30 GPa, PtH-II exhibits superconducting transition and \textit{T}$_{c}$ decreases with the applied pressure.
It is assumed that PtH-I does not show superconductivity above 4 K.
Remarkably, \textit{T}$_{c}$ (7 K at 30 GPa) of PtH-II is higher than that of powdered Pt by three orders of magnitude.
It is suggested that the hydrides of noble metals have higher \textit{T}$_{c}$ than the elements.

The starting material for the synthesis of PtH$_{\textit{x}}$ is a piece of Pt foil (Nilaco, 99.95 \%).
Previously, PtH$_{\textit{x}}$ (\textit{x}$\sim$1) was synthesized by compressing Pt/H$_{2}$ mixture using a DAC to a pressure of 25 GPa at room temperature \cite{Scheler2011,Hirao2008_1}. 
However, the electrical resistance measurements in the fluid or solid H$_{2}$ pressure transmitting medium is extremely challenging.
Since H$_{2}$ is highly compressible, the sample chamber significantly shrinks under compression, destroying the sample and electrodes. 
In this study, instead of pure H$_{2}$, we employed borane-ammonia complex NH$_{3}$BH$_{3}$ (SIGMA-ALDRICH, 97 \%) as a H$_{2}$ source.
At 1 bar, NH$_{3}$BH$_{3}$ releases H$_{2}$ on heating, and it ultimately decomposes to cubic boron nitride (\textit{c}-BN) and H$_{2}$ at 1000 K. 
We found that NH$_{3}$BH$_{3}$ releases H$_{2}$ even at pressures as high as 40 GPa by heating to a sufficiently high temperature above 1800 K. 
In this study, a Pt foil was confined with NH$_{3}$BH$_{3}$ powders in a sample chamber of DAC.
In the electrical resistance measurements, the electrodes (Au) and NaCl thin plate (thermal insulation between electrodes and a diamond anvil) were compressed together with the sample (The details of the sample configuration in a DAC are shown in sFig.1 in the Supplemental Material).
The sample and NH$_{3}$BH$_{3}$ were heated at a pressure exceeding 35 GPa by focusing an infrared (IR) laser ($\lambda$ = 1070 nm, SPI laser) on the sample Pt.
Keeping the temperature as high as 1800 K, the reaction of Pt and H$_{2}$ was monitored by \textit{in-situ} XRD at SPring-8/BL10XU.
After confirming that the entire sample transformed into PtH-II, the sample was quenched to room temperature to proceed to ac susceptibility (sample $\sharp$1) or electrical resistance (samples $\sharp$2 and $\sharp$3) measurements. 
Note that the formations of platinum nitride (PtN, Zinc-blende structure), gold hydride (See sFig. 2 in the Supplemental Material), and Na$_{3}$Cl and NaCl$_{3}$ were not confirmed in the present XRD measurements \cite{Gregoryanz2004,Zhang2013a}.
The pressure was determined using the shift in the first-order Raman band spectra of the diamond anvil facing the sample, with a proposed calibration \cite{Akahama2004b}.
The indicated pressures were those obtained by averaging the values obtained at room temperature before and after each cooling and warming cycle. 
Therefore, the actual pressures at low temperatures were thought to be slightly different from the indicated values.
In the present report we address our sample PtH$_{\textit{x}}$ as the exact value of \textit{x} cannot be determined.

\begin{figure}
\includegraphics[scale=0.4]{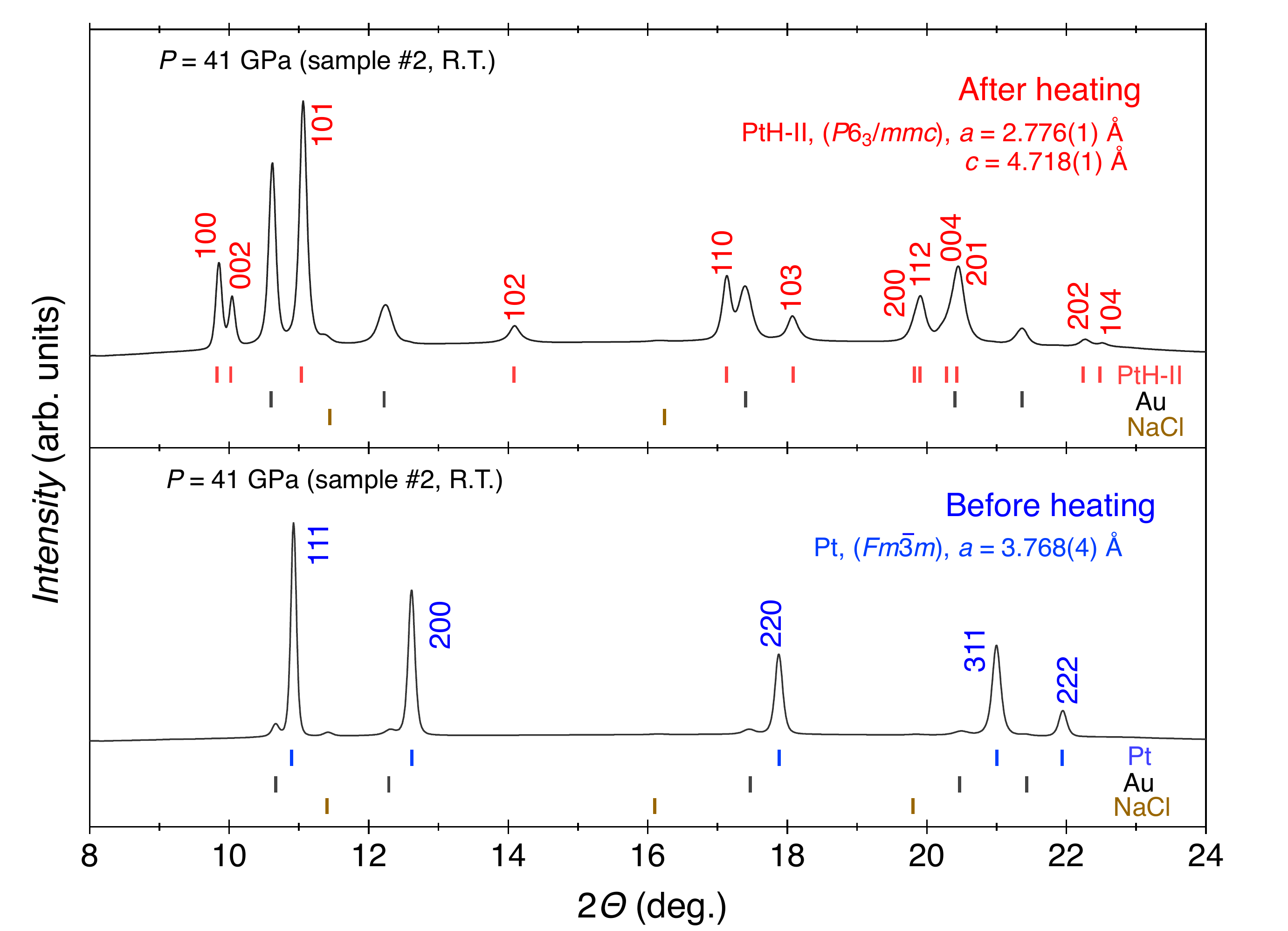}
\caption{\label{Figure 1} (color online) XRD profiles of the sample $\sharp$2 before and after the synthesis of PtH$_{\textit{x}}$. The wavelength of the monochromatic synchrotron X-ray beam was 0.4145 \AA. Vertical bars indicate the diffraction peak positions of PtH-II (\textit{P}6$_{3}$/\textit{mmc}), Pt (\textit{Fm}$\bar{3}$\textit{m}), Au (\textit{Fm}$\bar{3}$\textit{m}, electrodes), and NaCl (CsCl-type, thermal insulation) at 41 GPa. The smooth and broad backgrounds in the XRD profiles are from Compton scattering of a diamond anvil. The diffractions from NH$_{3}$BH$_{3}$ and \textit{c}-BN are weak and barely visible in the XRD profiles shown here.}
\end{figure}

Figure 1 shows the typical XRD profiles obtained for sample $\sharp$2 at room temperature before and after the laser heating.
The crystal structure and lattice parameters (\textit{P}6$_{3}$/\textit{mmc}, \textit{a} = 2.776(1) \AA, \textit{c} = 4.718(1) \AA  at 41 GPa) of our samples showed excellent agreements with that of PtH-II synthesized in high pressure H$_{2}$ at room temperature \cite{Scheler2011,Hirao2008_1}. 
Besides, our XRD measurements confirmed that PtH-II was stable at room temperature for at least 12 hours after its synthesis (Supplementary Material sFig. 3). 
In this study, we synthesized all three samples (sample $\sharp$1 $\sim$ $\sharp$3) using the same method.

Figure 2 shows the temperature dependence of the real part of the ac magnetic susceptibility ($\chi$$\prime$) of PtH$_{\textit{x}}$ (sample $\sharp$1) at the given pressures.  
We performed the measurements upon pressure decrease at 36 GPa, 32 GPa, and 28 GPa. 
The sharp drop of the real part in ac susceptibility ($\chi$$\prime$) appeared at 36 GPa and 32 GPa at 4.8 K and 6.7 K, respectively. 
Using the analysis discussed in the Supplementary Material, the observed $\Delta$$\chi$$\prime$ = 6 - 11 nV jump at \textit{T}$_{c}$ is consistent with perfect diamagnetism, the hallmark of bulk superconductivity. 
At 28 GPa, near the pressure where PtH$_{\textit{x}}$ transforms from PtH-II to PtH-I upon pressure decrease, no diamagnetic signal is visible above 4.6 K \cite{Scheler2011}. 
It is assumed that PtH-I constitutes the majority  of sample $\sharp$1 at 28 GPa, and the superconducting transition vanishes at temperatures above 4.6 K.
We conclude that the observed superconducting transitions at 32 GPa and 36 GPa are from PtH-II.
\begin{figure}
\includegraphics[scale=0.4]{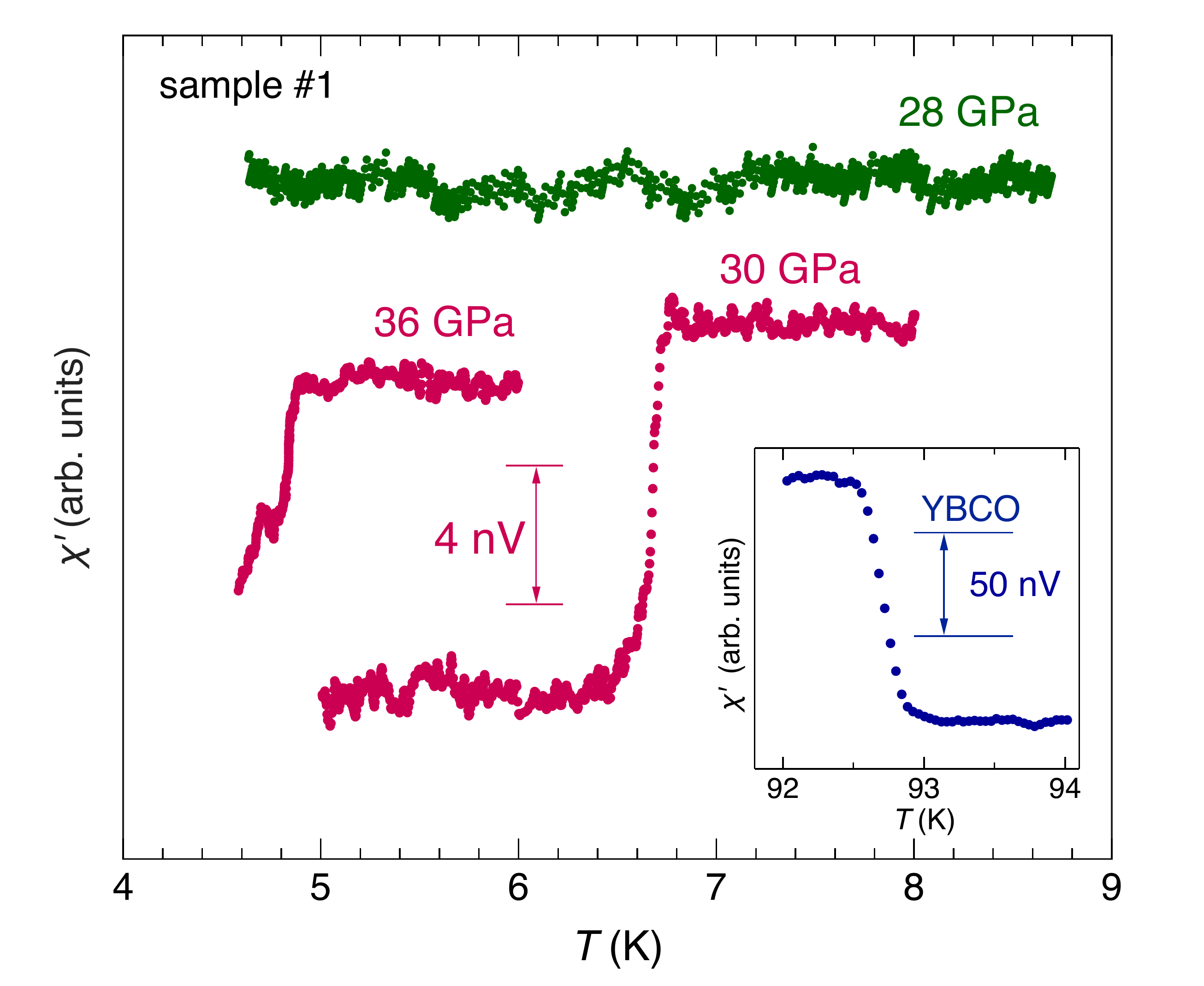}
\caption{\label{Figure 2} (color online) The $\chi$$\prime$ vs. \textit{T} for PtH$_{\textit{x}}$ as pressure is decreased from 36 to 28 GPa. The superconducting transition shifts under decreasing pressure to higher temperatures. No superconducting transition is observed at 28 GPa. \textit{T}$_{c}$ is determined from the temperature at the transition onset. All data are on the same scale but shifted vertically. The superconducting transition signal of YBa$_{2}$Cu$_{3}$O$_{7-\textit{x}}$ (YBCO) appears at about 92.6 K in the direction opposite to the signal from PtH$_{\textit{x}}$ (See also the Supplementary Material).}
\end{figure}

To obtain further evidence of superconductivity, we performed the electrical resistance measurement of PtH-II (sample $\sharp$2 and sample $\sharp$3).
Figures 3a and 3b show the temperature dependence of electrical resistance (\textit{R}) of PtH-II at 35 GPa (sample $\sharp$2). 
Note that the electrical resistance measurements were performed using a two-point contact method due to technical difficulties.
Each Au electrode was split into two wires in the vicinity of the samples (See sFig. 1 in Supplementary Material).
In addition, in the experiments of sample $\sharp$3, one of the four electrodes (V2 in sFig 1b) was broken, and the electrodes were re-connected outside a DAC using Cu wires. 
The measured \textit{R} includes the contact resistance between electrodes and samples and the resistance of the Cu wires (for sample$\sharp$3).
Consequently, the \textit{R} of samples $\sharp$2 and 3 does not drop to zero at \textit{T}$_{c}$. 
In Fig. 3a, the \textit{R} vs. \textit{T} curve has a positive slope at temperatures below 300 K, clearly indicating that PtH-II is a metal. 
At 7.5 K, \textit{R} exhibits a sharp decrease deviating from linear dependence to temperature. 
To investigate the origin of the reduction in \textit{R}, we performed the electrical resistance measurements on sample $\sharp$3 at 35 GPa in an external magnetic field (Fig. 3b). 
In the zero-field-cooled (ZFC) measurement at 35 GPa, \textit{R} shows a sharp drop at 7 K in agreement with sample $\sharp$2. 
When the magnetic field \textit{H} = 0.1 T is applied, the decline in \textit{R} is at 3.5 K, and it vanishes for temperatures above 2.8 K at \textit{H} = 0.5 T.
The shift in \textit{T}$_{c}$ to lower temperatures in an external magnetic field is one of the characteristics of superconductivity.
The observed shift in resistance dropping temperature strongly suggests the superconducting transition of PtH-II.
It is noted that the critical magnetic field estimated by using the data of electrical resistance measurements in the magnetic field is less than 0.5 T. 
Since the DAC and gasket are not perfectly nonmagnetic at low temperatures, the actual applied magnetic flux density might be bigger than the indicated value resulting in the significant suppress of \textit{T}$_{c}$.

Gathering the results, these experiments provide unambiguous evidence for the superconductivity of PtH-II.
Remarkably, the observed \textit{T}$_{c}$ of PtH-II is three orders of magnitude larger than that of compacted powder Pt \cite{Konig1999}. 

\begin{figure}
\includegraphics[scale=0.4]{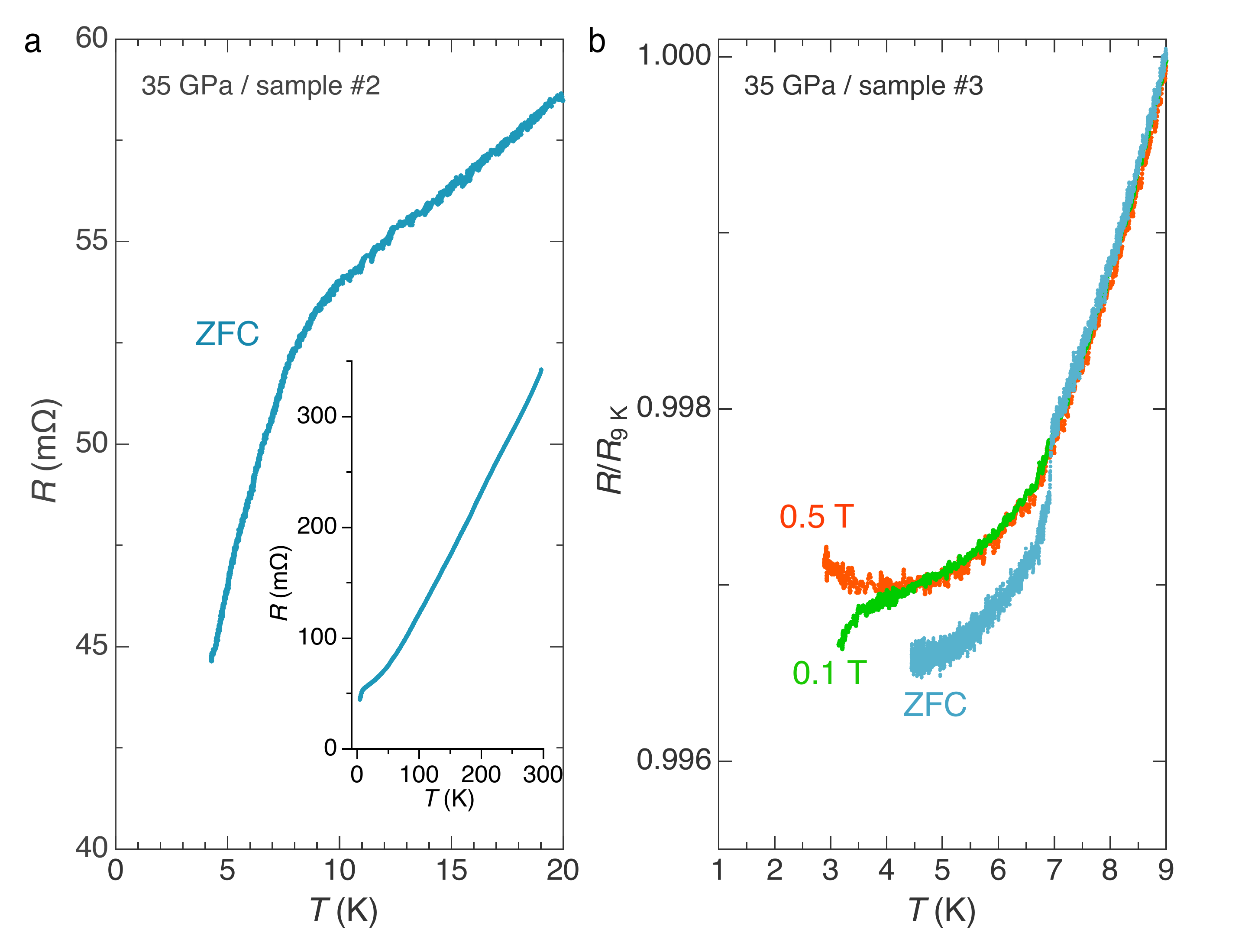}
\caption{\label{Figure 3} (color online) The \textit{R} vs. \textit{T} of PtH-II at 35 GPa in external magnetic fields. $\bf{a}$ \textit{R} vs. \textit{T} curve of PtH-II (sample $\sharp$2) at 35 GPa in ZFC. The inset graph indicates the data from 4.2 K to 300 K. $\bf{b}$ \textit{R} vs. \textit{T} curves of PtH-II (sample $\sharp$3) at 35 GPa in 0 T (ZFC), 0.1 T, and 0.5 T field-cooled. The \textit{R} data were scaled at 9 K for comparison.}
\end{figure}

For discussion, we compare the experimental results with theoretical predictions. 
In Fig. 4, the \textit{T}$_{c}$ of sample $\sharp$1 (ac magnetic susceptibility measurements) is plotted as a function of pressure along with the predicted \textit{T}$_{c}$s and structural phase diagrams  \cite{Scheler2011,Kim2011,Hirao2008_1}. 
Since the pressures at \textit{T}$_{c}$s could not be measured, we have not plotted the \textit{T}$_{c}$s obtained for different samples ($\sharp$1 $\sim$ 3) in the same graph. 
The present experimental observation and theoretical predictions agree with each other at the points where PtH-II is a superconductor and \textit{T}$_{c}$ decreases with pressure. 
According to the \textit{ab-initio} calculations, the negative d\textit{T}$_{c}$/d\textit{P} slope originates from the monotonical decrease in \textit{e}-\textit{p} coupling constant and the saturation of averaged logarithmic phonon frequency with increasing pressure \cite{Scheler2011,Kim2011}. 
The observed \textit{T}$_{c}$s are almost half of the predictions based on a harmonic approximation and much higher than 0.4 K which has been suggested by calculations that consider anharmonic effects \cite{Errea2014}. 
Although the present results do not clearly answer the question about the anharmonic effect on \textit{T}$_{c}$, it is assumed that anharmonicity of a phonon does not significantly suppress the \textit{T}$_{c}$ in PtH-II. 
In this study, we could not determine the stoichiometry of our samples.
So far, it has been observed that the \textit{T}$_{c}$ of PdH$_{\textit{x}}$ increases proportionally to the hydrogen content \textit{x} = H/Pd in PdH$_\textit{{x}}$ \cite{Skoskiewicz1972,Stritzker1972}. 
If the observed \textit{T}$_{c}$ is for PtH$_{\textit{x}}$ with \textit{x} far less than 1, there would be a better probability of obtaining a higher \textit{T}$_{c}$ by improving the stoichiometry. 
Further studies using a sample with a perfect stoichiometry PtH would provide more concrete results and answer the question about anharmonicity.

\begin{figure}
\includegraphics[scale=0.35]{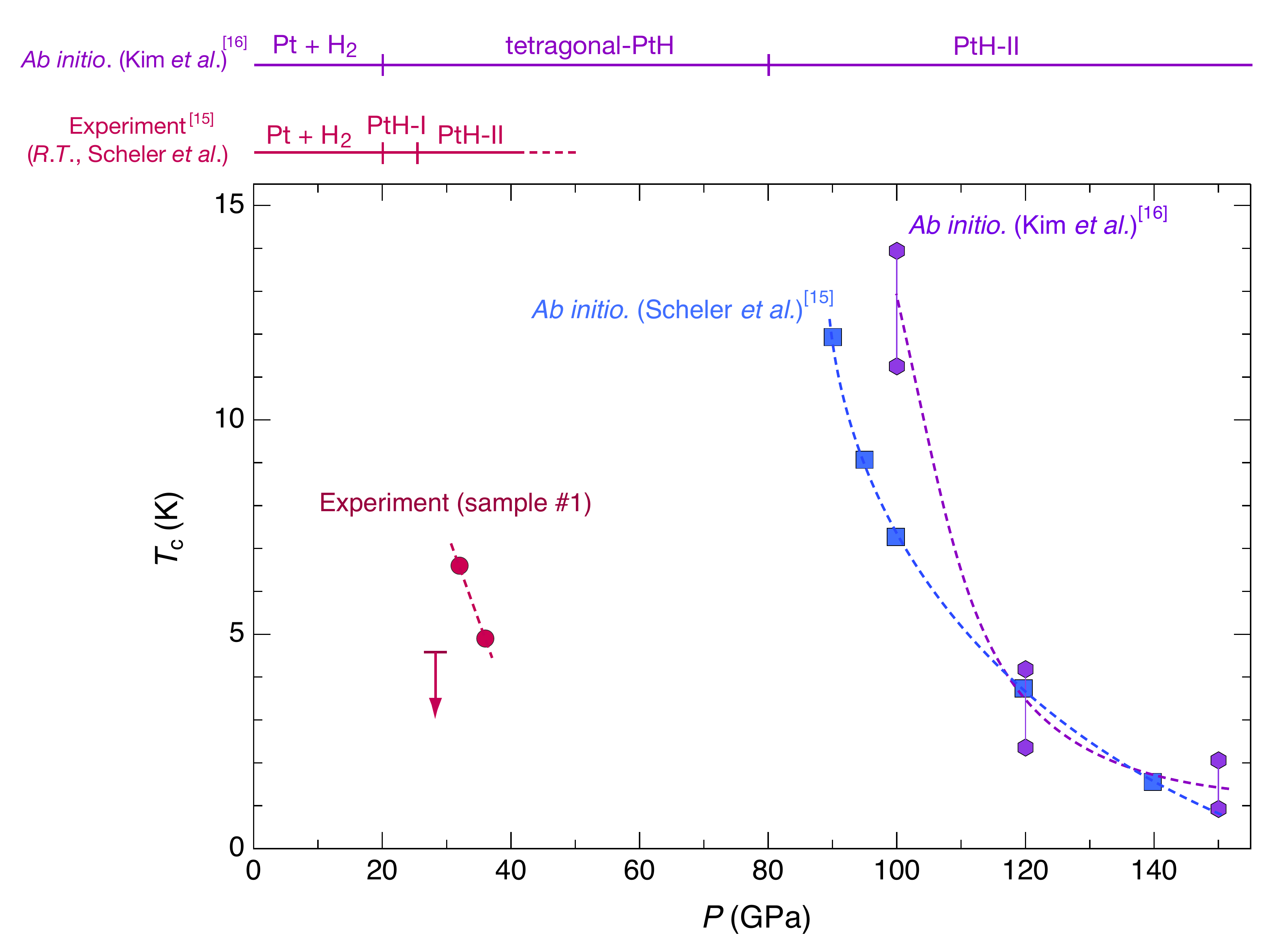}
\caption{\label{Figure 4} (color online) The \textit{T}$_{c}$ vs. \textit{P} for PtH$_{\textit{x}}$. The structural phase boundaries observed by the experiments\cite{Scheler2011} and predicted by an \textit{ab-initio} calculation\cite{Kim2011} are indicated above the graph. The \textit{ab-initio} calculations of Refs. 15 and 16 predict that PtH-II is stable at pressures above 80 GPa. The down arrow at 28 GPa indicates that the superconducting transition is not observed above 4.6 K. The broken lines overlapping  \textit{T}$_{c}$s are the guides for eyes to indicate the pressure dependence of \textit{T}$_{c}$s.}
\end{figure}

In most superconducting \textit{d}-metals, the presence of hydrogen in the metal lattice either decreases the \textit{T}$_{c}$ (e.g., TcH, ReH$_{x}$, VH, NbH, ZrD$_{x}$) or does not affect it (e.g., RuH, MoH) \cite{Antonov1987, Bashkin2000, I.O.BashkinV.E.2003}. 
At present, PdH and PtH are the only material hydrides that have higher \textit{T}$_{c}$ than elemental metals \cite{Antonov2002b, Donnerer2013, Burtovyy2004, Scheler2013, Kalita2010, Zaleski-Ejgierd2012, Huiberts1996, Skoskiewicz1972,Stritzker1972}. 
The \textit{ab-initio} calculation performed by Kim \textit{et al.} predicted the formation of AuH, silver hydride (AgH), rhodium hydride (RhH), and iridium hydride (IrH)$_{8}$ under high pressure. 
For all of them, except AgH, superconductivity has been predicted \cite{Kim2011}. 
The hydrides of these noble metals may possibly exhibit superconductivity at a higher temperature than the elements.
Although RhH does not exhibit superconducting transition down to 0.3 K at ambient pressure, it would be worth extending the search to high pressures \cite{Antonov1987}. 

In summary, we have revealed the superconducting transition of PtH-II by the ac-susceptibility, electrical resistance, and XRD measurements under high pressure.
The \textit{T}$_{c}$ of PtH-II is 6.7 K at 32 GPa and decreases with applied pressure.
The results provide a definitive answer to the question whether a superconductor from Pt and H can be developed. 
Similarly to the case of PdH, the observed highest \textit{T}$_{c}$ of PtH-II is more than 10$^3$ times higher than that of powder Pt \cite{Konig1999,Skoskiewicz1972,Stritzker1972}.
The observation of superconductivity in PtH-II stimulates the interests in other noble metal hydrides because they possibly form a group whose metal hydrides have higher \textit{T}$_{c}$ than that of elemental metals.
Extending the search for superconductivity to other noble metal hydrides, including the observation of the isotope effect on the \textit{T}$_{c}$ of deuteride samples, would advance our understanding of superconductivity in metal hydrides and indicate methods to obtain new superconductors.
Finally, the present experiments are significant in the field of high-pressure science. 
Pt has been used in high-pressure experiments as a pressure calibrant, electrical probe material, and a laser absorber both at high and low temperatures.
With growing interests in high-temperature superconducting metal hydrides, our results advice against the use of Pt as an electrode that will be in contact with hydrogen and hydrides.

\begin{acknowledgments}
We thank Prof. James S. Schilling and Prof. James J. Hamlin for guiding us in the ac-susceptibility measurements using DACs. We appreciate the help by Mr. Kazuki Kawai in ac-susceptibility measurements. YBCO samples used for ac-susceptibility measurements were provided by Prof. Hidekazu Mukuda and Prof. Mitsuharu Yashima. The laser heating and XRD measurements were performed at BL10XU/SPring-8 (Proposal Nos. 2015A1256, 2015B35, and 2017B1492) and Nagoya University with the help of Prof. Ken Niwa and Prof. Masashi Hasegawa. This study was supported by JSPS KAKENHI Grant Number 25800195, Specially Promoted Research (26000006), and Nikki-Saneyoshi (JGC-S) Scholar Ship Foundation Grant for Young Researchers.
\end{acknowledgments}

\providecommand{\noopsort}[1]{}\providecommand{\singleletter}[1]{#1}%

\end{document}